\documentstyle[prl,twocolumn,aps,epsf,epsfig]{revtex}
\begin{document}
\draft
\twocolumn[\hsize\textwidth\columnwidth\hsize\csname@twocolumnfalse%
\endcsname

\title{Interlayer Magnetic Coupling and the Quantum Hall Effect in
Multilayer Electron Systems.}

\author{
Luis Brey 
}

\address{Instituto de Ciencia de Materiales (CSIC).
Campus de la Universidad Aut\'onoma,
28049, Madrid, Spain.  }

\date{\today}

\maketitle

\begin{abstract}
\baselineskip=2.5ex
We study the effect that 
the electron-electron interaction 
has on the properties of a multilayer electron system. We consider the case 
corresponding to filling factor unity in each layer.
We find that as a function of the sample parameters the system has 
ferromagnetic, canted antiferromagnetic or 
paramagnetic interlayer spin correlations.
These three ground states are QHE phases, because
of the existence of a finite activation energy.
In the ferromagnetic phase the gap is due to the intrawell exchange energy, 
whereas in the paramagnetic phase the gap appears due to the spatial
modulation of the interwell coherence.
\end{abstract}

PACS number 73.40.Hm
]


The quantum Hall effect (QHE) is one of the most striking phenomena observed
in two dimensional electron gas (2DEG) systems\cite{prange}. 
The QHE occurs because the 2DEG becomes incompressible 
at certain filling  factors.  
In the fractional QHE, the energy gap source of the  incompressibility
is produced  by interactions between electrons.
In the  even integer QHE the incompressibility   
is due to the quantization of the electron kinetic energy and  interactions
are not important for the occurrence of the QHE. 
In the case of the  odd integer QHE the incompressibility 
occurs because at odd filling factors    the 2DEG is
ferromagnetic. Due to the electron-electron interaction 
the ground state of a 2DEG at odd filling , is completely spin polarized even
for zero Zeeman energy\cite{sondhi}.       
Given the new physics which appears in 2DEG in the QHE regime, 
the question that arises is whether the quantum Hall phases are unique
to 2DEG or  they can occur in three dimensional (3D) 
conductors\cite{BIH1}.
In this direction, some studies in 
narrow gap 3D semiconductors in the strong  magnetic field  limit have shown 
some
signatures of an incipient quantum Hall phase\cite{murzin}.
On the  other hand  
the progress   in epitaxial growth has made possible to fabricate
semiconductor systems where 2DEG's  with 
extra degrees of   freedom  exists: 
\par \noindent
i) Wide parabolic quantum wells where a thick  
electron gas layer ($\sim$2000$\AA$) is formed. This   system present
a  clear QHE phase\cite{wpqw}.
\par \noindent
ii)Double layer 2DEG systems with electrons confined to two
parallel sheets separated by a distance comparable to that between
electrons within a plane\cite{dqw1}. Double layer systems 
present QHE at total integer filling factors, even in absence of
tunneling between the electron planes.
\par \noindent
iii) Superlattices, where an appreciable dispersion 
of the electronic spectrum
in the direction perpendicular  to the layers exits.
Accurately quantized Hall plateaus have been  observed  in these
multilayer systems\cite{stormer,druist}
when a magnetic field is applied parallel to the superlattice axis.
Studies of vertical transport
in these supperlattices have shown\cite{druist} the existence
of a chiral two-dimensional system that form at the surface o the layered
system\cite{edget}.
\par \noindent
The poor mobility of the bulk narrow gap semiconductors and
the small number of new degrees of freedom of
the parabolic   and  double quantum wells with respect the  2DEG,  made 
the superlattices  the best  candidates for 
studying QHE phases in 3D conductors.

\vspace{-5cm}
\begin{figure}
\epsfig{figure=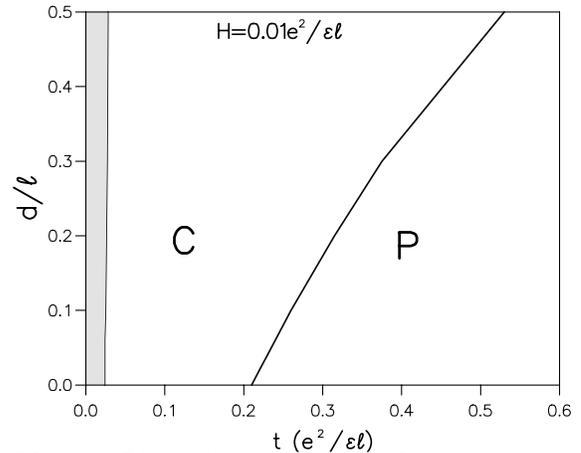,width=8.5cm}
\caption{
Phase diagram for a multilayer system, with filling  factor unity in each 
well. The Zeeman coupling is $H$=0.01$e^2/\epsilon \ell$ and
the electron layer thickness $b=0.8\ell$. Three
phases are present: a ferromagnetic phase (shadow region),
a canted antiferromagnetic region (C) and a paramagnetic region (P).
}
\end{figure}
In this work we study the effect that  the electron electron interaction
has on the properties of the multilayer electron system. We consider the 
case corresponding to filling factor unity in each electron layer.
Two  points are raised in this paper: the magnetic order of the 
electron layers and the conditions for the occurrence of the QHE in this
system.
The main results we obtain are the following:
i)As a function of the sample parameters (Zeeman coupling, $H$,
interlayer tunneling, $t$, and barrier thickness, $d$)
we obtain that the system changes from a QHE state with interlayer
ferromagnetic
spin correlations to a new QHE state with canted antiferromagnetic interlayer
correlations (see Fig.1). For larger values of the tunneling amplitude, 
we obtain that the
system undergoes another phase transition towards a paramagnetic state.
These transitions are second order phase transitions.
ii)We also study the value of the activation energy as a function of the 
sample parameters. In Fig.2 we plot this energy gap 
\vspace{-5cm}
\begin{figure}
\epsfig{figure=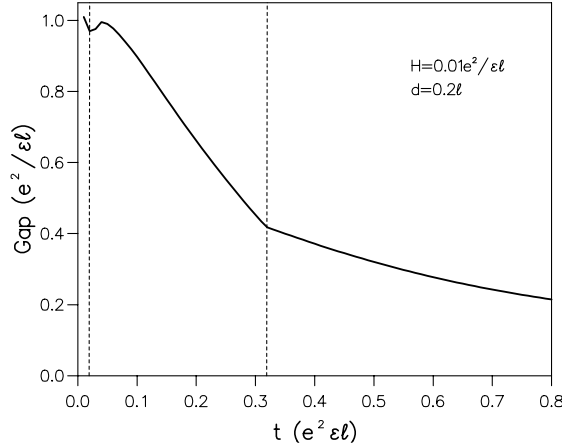,width=8.5cm}
\caption{
Activation energy of a multilayer system as a function of the
tunneling amplitude.
The Zeeman coupling is $H$=0.01$e^2/\epsilon \ell$,  the barrier thickness
$d=0.2 \ell$ and the electron layer thickness is $b=0.8\ell$.
The vertical  lines indicate the values of $t$ where the different
phase transitions occur.
}
\end{figure}

\par \noindent
as a function of the
tunneling amplitude for a multilayer system with $d=0.2 \ell$ and
$H = 0.01 e ^ 2 / \epsilon \ell$ (here $\ell$ is the magnetic length).
We find that this gap is finite even for very large values of $t$, where
the system is paramagnetic. This implies that the paramagnetic phase of
the multilayer system is also a QHE phase. As we explain below, the energy
gap in the paramagnetic phase appears because the system breaks 
spontaneously the translational symmetry along the multilayer axis
by modulating the interwell coherence.
From our results  we conclude that in multilayer electron systems,
with filling factor unity in each well, the QHE prevails in all phases
due to the existence of a finite activation energy, even  at $d,t  \rightarrow
\infty$.

We treat the electron electron interaction in the   Hartree-Fock (HF) 
approximation.
In 2DEG's  in the QHE regime, this  approximation\cite{kallin}
gives results which agree well with experiments\cite{pinczuk}. In double
quantum well systems, at total filling factor unity, the HF approximation
is less accurate because of the competition   between inter and intrawell
correlations,  nevertheless for small values    of the layer separation
the agreement  with experiments is rather good\cite{dqw1}.
In double layer systems at total filling factor  two, the HF approximation
gives reasonable results because of the existence of filled  Landau levels
and the existence of  charge   excitation gaps, which make the HF ground state
to be a   good approximation to the real many-body ground state\cite{zheng}.
From the above, the HF approximation is expected to be a good approximation for
describing weakly coupled multilayer systems at filling 
factor unity in each well.
For stronger coupled multilayer systems we expect  HF 
to be a reasonable approximation for the real many-body ground state, because
the existence of filled bands and charge excitation gaps.

The calculations  presented here employ realistic Coulomb interaction
potentials, take into account       interlayer tunneling and Zeeman 
coupling and therefore we expect our results to be qualitatively
and quantitatively trustworthy. 
We take the multilayer vertical axis as the $z$-direction and the electrons live
in the $x-y$ planes. 
The magnetic field, $B$, is applied in the $z$-direction.
Since  $B$ is very strong, we
only consider states in the
lowest  energy Landau level of the   lowest energy
subband of each well.  The Hamiltonian of the system is written   as
$\hat H =H_0 + V$, with
\begin{eqnarray}
H_0  = & -& t \sum _{<i,j>,k,\sigma} \left ( C ^   + _{i,\sigma,k}
C_{j,\sigma,k} + h.c. \right ) \nonumber \\
& - & H  \sum  _{i,\sigma,k} \sigma    
C^+ _{i,\sigma,k} C_{i,\sigma,k} \,  \, \, \, ,
\end{eqnarray}
where $C^+ _{i,\sigma.k}$ creates an electron in the lowest Landau
level in layer $i$ with spin $\sigma$ ($\sigma = \pm   1$) in the direction
of $B$, and with intra   Landau level index $k$. The sum in the first
term of Eq.1 is over first     neighbors layers.
The many-body    part    of $\hat H$ takes the form,
\begin{eqnarray}
V =  & & \! \! \! {1 \over  {2S}} \sum _{\sigma ,  \sigma '}
\sum  _{i,j} \sum_ {k,k',{\bf q}}
V_{i,j} (q) e ^{- q ^2  \ell ^2 /2} e ^{ i q_x (k -k')   \ell ^2}
\nonumber    \\
& \times    & C^+ _{i,\sigma,k+q_y} C^+_{j,\sigma ',k'}
C_{j,\sigma ' ,k' +q_y} C_{i,\sigma,k}
\, \, \, ,
\end{eqnarray}
where   $S$ is the sample area
and the interaction potential has  the   form 
$V_{i,j} =   2 \pi   e ^2 / \epsilon q \, F_{i,j} (q,b,d)$, 
with 
$F_{i,j}$ being finite layer thickness form    factors\cite{cote}, which
depends on the thickness of the 2DEG in each layer, $b$, and on the     
barrier thickness, d.

The  multilayer systems are doped in the barriers and in absence of $B$,
all layers have the 
same number of electrons. 
We consider that this is 
also the situation at large values   of $B$,
since any other
situation it would cost a large Hartree energy. 
At filling factor unity in each layer   the interwell correlation is
not very important  so we only consider solutions with translational
symmetry in the plane ($x,y$) of the electron gases.
Broken   translational symmetries along the multilayer axis ($z$-direction)
are allowed, always with the condition of having filling factor unity in each layer.
With these constrains, the HF expectation value of $V$ takes the form,
\begin{eqnarray}
<V>=  &-& {1 \over {2   S}} \sum  _{i,j,{\bf q}} 
\sum _ {\sigma,\sigma ' ,k }
V_{i,j} (q) e ^  {- q ^2  \ell ^2 /2}   \nonumber \\ & &  \!  \! \! \! \! 
<C^+ _{i,\sigma, k} C _{j,\sigma ' ,k} >
<C^+ _{j,\sigma ' ,k-q_y} C_{i,\sigma, k-q_y} > 
\, \, \, ,
\end{eqnarray}
here the sum in $k$  is over all its  possible  values. 
By minimizing the energy $< \hat H >=<H_0>+<V>$, we obtain the energy of the
ground state
of the system and its properties. 

We have solved the Hamiltonian for different values of $d$, $t$ and
$H$. For each layer $i$ we calculate the expectation value of the
total spin operator per electron $<{\bf S} _i>$, 
\begin{eqnarray}
& & S_{i,z}= {1 \over {N_\phi}} \sum _{k, \sigma} \sigma
C^+ _{i,\sigma, k} C _{i,\sigma,k} \nonumber \\
& &  S_{i,x}+iS_{i,y} = {1 \over {N_\phi}} \sum _{k}
C^+ _{i,+1, k} C _{i,-1,k} 
\, \, \, ,
\end{eqnarray}
here $N_ \phi = S / 2 \pi \ell ^ 2$ is the the Landau level degeneracy.
For characterizing the   ground state it is also necessary 
to quantify the  interlayer coherence which is given by 
the following expectation value,
\begin{equation}
\Delta _{\sigma, \sigma '} (i)  =  {1 \over {N _ \phi}}
\sum _ k 
< C^ + _{  i , \sigma ,k} C _{  i +1 , \sigma ,k}> \, \, \, . 
\end{equation}
This quantity  represents the coherence between wells. 
\vspace{-5cm}
\begin{figure}
\epsfig{figure=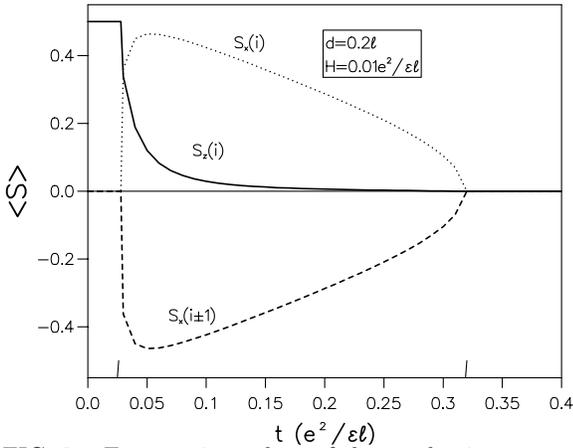,width=8.5cm}
\caption{
Expectation values of the total spin operator per electron
as a function of the tunneling amplitude.
The Zeeman coupling is $H$=0.01$e^2/\epsilon \ell$, the barrier thickness
$d=0.2 \ell$ and the electron layer thickness is $d=0.8\ell$.
The vertical  tickmarks in the lower $x$ axis  
indicate the values of $t$ where the different
phase transitions occur.
}
\end{figure}

Looking to the values
of $<{\bf S} _i>$ we find three different classes of
ground states (see Fig.1 and Fig.3):
\par \noindent
1){\it Ferromagnetic} phase, where all electron layers are fully spin
polarized in the direction of the magnetic field, i.e.
$<{\bf S} _i> = (0,0,1/2)$. This phase occurs for small values of $t$
or large values of $H$. In the ferromagnetic phase the intralayer 
coherence is more important than the interlayer coherence and all the
expectation values of the operators
$\Delta _{\sigma , \sigma '} (i)$ are zero,
and there is not vertical kinetic energy contribution to the total energy.
The ferromagnetic ground state is a QHE phase, and the activation gap  
(see Fig.2) is the cost in energy of adding 
an electron to the system with the spin pointing antiparallel 
to the magnetic field.
In this phase the ground state has the same translational 
symmetry than  the original Hamiltonian.
\par \noindent
2){\it Canted antiferromagnetic} phase. In this phase the total spin in 
each layer acquires a component perpendicular to the magnetic field,
$<{\bf S} _i>= ({\bf S}_{i, \perp}, S_{i,z})$, and the magnitude of
$<{\bf S} _i>$ is smaller than its maximum value, $1/2$. In this phase
the sign of ${\bf S}_{i, \perp}$ alternates from layer to layer, i.e.
${\bf S}_{i, \perp}= -{\bf S}_{i \pm 1, \perp}$, so that the translational
symmetry of the Hamiltonian is spontaneally broken and the unit cell in the
$z$-direction consists of two electron layers which are labelled 
$2i-1$ and $2i$. The $z$-component of
$<{\bf S} _i>$ is finite and it has the same value in all layers, therefore
in the canted phase there is an interlayer antiferromagnetic coupling of the
transverse component of the total spin\cite{comm}.
By performing calculations in bigger size unit cells, we have checked
that this phase is stable with respect spiral ordering
of the transverse component of $<{\bf S} _i>$.

\vspace{-5cm}
\begin{figure}
\epsfig{figure=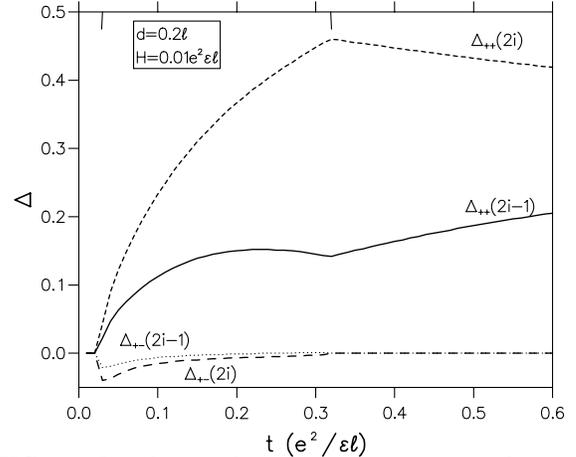,width=8.5cm}
\caption{
Interlayer coherence parameters as a function of the tunneling amplitude.
$\Delta _{++} (i)$ are real and we plot its real part.
$\Delta _{+-} (i)$ are imaginary  and we plot its imaginary part.
The Zeeman coupling is $H$=0.01$e^2/\epsilon \ell$, the barrier thickness
$d=0.2 \ell$ and the electron layer thickness is $b=0.8\ell$.
The vertical dashed tickmarks in the upper $x$ axis  
indicate the values of $t$ where the different
phase transitions occur.
}
\end{figure}
In the canted phase the interlayer coherence parameter
is  different from zero and 
verify the relations:
\begin{eqnarray}
\Delta _{+,+} (i) & =&  \Delta _{-,-} (i) 
\nonumber \\
\Delta _{+,-} (i)&  = &  - \Delta _{-,+}  ^ * (i)\, \, \, , 
\end{eqnarray}
but  the interlayer coherence parameter
depends on $i$.
In Fig.4 we plot 
$\Delta _{+,+} (2i)$,
$\Delta _{+,+} (2i-1)$,
$\Delta _{+,-} (2i)$ and 
$\Delta _{+,-} (2i-1)$,
as a function of $t$.
We see that \begin{equation}
\Delta _{\sigma, \sigma '}(2i) \neq 
\Delta  _{\sigma, \sigma '} (2i-1)\, \, \, \, ,
\nonumber \end{equation}
and this implies that there is a modulation of the interlayer coherence. 
Therefore in the canted phase the translational symmetry along the 
multilayer axis is broken not just by the antiferromagnetic ordering of the
layers, but also by the modulation of the interlayer coherence.
In the selfenergy calculation, the modulation of the interlayer coherence acts
as a spatial modulation of the hopping amplitude, and
this modulation contributes to the opening of 
an energy gap at the Fermi level.
The canted  phase appears at intermediate values of the tunneling amplitude,
see Fig.1 and Fig.3, and the reason for its existence is that in this phase the
system can take advantage of the kinetic energy by creating interlayer 
antiferromagnetic spin correlations. The antiferromagnetic order is canted in
order to minimize the lost of Zeeman energy.
The canted ground state is a QHE phase. The transport
activation energy is finite 
because the system is partially spin polarized in the direction of $B$,
and 
because the interlayer coherence is spatially modulated.

Canted ground states corresponding to rotations of all the
${\bf S}_{i, \perp}$ are degenerated and therefore this phase should get
a gapless collective mode associated with this degeneracy.

\par \noindent
3){\it Paramagnetic phase} In this phase the expectation value of the 
total spin operator
is zero in all  layers, $<{\bf S} _i > =0$. 
This phase occurs at large values of $t$, where the kinetic energy
and interwell coherence energy is much bigger than the Zeeman and 
intrawell exchange energy. In this phase
$\Delta _{+,-}(i)=0$ but the equal spin interwell coherence
parameters are different from zero and verify,
$\Delta _{+,+} (2i-1)$=$ \Delta _{-,-}(2i-1) 
$$\neq$$ \Delta _{+,+}(i)$=$\Delta _{-,-}(2i)$. 
In this phase, the system breaks spontaneously the translational symmetry
by modulating the interwell coherence along the vertical axis of the multilayer.
In the paramagnetic ground state the unit cell consists of two electron layers.
This modulation of the interwell coherence creates an energy gap at the
Fermi energy, and the paramagnetic ground state is a QHE phase.

The superlattices studied in references \cite{stormer,druist} have
thick barriers and they are in the ferromagnetic phase. For studying
antiferromagnetic and paramagnetic phases, it is necessary multilayers
with thin barriers and large tunneling amplitudes. The multilayers
are usually doped in the barriers and in the case of thin barriers this 
produces a strong scattering of the electrons by impurities, which prevents
many-body driven ground states. It is possible to circumvent this problem
by working with superlattices superimposed on wide parabolic wells.
\cite{breysl,santos,sundaram}. These systems are remotely doped and it is 
possible to obtain multilayers with thin barriers and high electron mobility.
The ground states magnetic properties can be studied experimentally
by using optically pumped nuclear magnetic resonance. This technique has been
very useful for the study of the magnetic nature of 2DEG's\cite{tycko}.
Also it could be very useful the application of a 
magnetic field, $B_{\parallel}$,
parallel to the electron sheets. 
$B_{\parallel}$ changes the value of the Zeeman coupling, and as 
a function of its strength the ground state of the system could change.
This phase transition could be identified by studying the activation 
energy as a function of $B_{\parallel}$\cite{eisenstein}.
Strong enough $B_{\parallel}$ also destroys the interlayer 
coherence\cite{murphy,yang,cote1}. In the paramagnetic phase the activation 
energy is due to the spatial modulation of the interlayer coherence, 
and the application of a strong $B_{\parallel}$, would destroy the QHE.

In conclusion, we have studied the effect that 
the electron-electron interaction 
has on the properties of a multilayer electron system. We consider the case 
corresponding to filling factor unity in each layer.
We have found that as a function of the sample parameters the system has 
ferromagnetic, canted antiferromagnetic or 
paramagnetic interlayer spin correlations.
We have obtained that these three ground states are QHE phases, because
of the existence of a finite activation energy.
In the ferromagnetic phase the gap is due to the intrawell exchange energy, 
whereas in the paramagnetic phase the gap appears due to  the spatial
modulation of the interwell coherence.

This work was supported by the CICyT of Spain under contract N. PB96-0085, and
by the Fundaci\'on Ram\'on Areces. Helpful conversations with C.Tejedor 
and Luis Mart\'{\i}n Moreno
are grateful acknowledged.


\begin{references}
\bibitem{comm}This phase is similar to that found in reference [12],
for 
double quantum wells
at total filling factor 2.
\end{references}
\end{document}